\documentclass[a4paper,11pt]{article}
\usepackage{pos}

\title{Simple BSM explanations of $\amu$ in light of the FNAL muon $g-2$ measurement}

\author[a,b]{Peter Athron}
\author[b]{Csaba Bal\'azs}
\author*[b]{Douglas Jacob}
\author[c]{Wojciech Kotlarski}
\author[c]{Dominik St\"ockinger}
\author[c]{Hyejung St\"ockinger-Kim}

\affiliation[a]{Department of Physics and Institute of Theoretical Physics, Nanjing Normal University,\\
	Nanjing, Jiangsu 210023, China}

\affiliation[b]{School of Physics and Astronomy, Monash University,\\
	Melbourne, Australia}

\affiliation[c]{Institut f\"ur Kern- und Teilchenphysik, TU Dresden,\\
	Zellescher Weg 19, Dresden, Germany}

\emailAdd{douglas.jacob@monash.edu}
\emailAdd{peter.athron@coepp.org.au}
\emailAdd{csaba.balazs@monash.edu}
\emailAdd{wojciech.kotlarski@tu-dresden.de}
\emailAdd{Dominik.Stoeckinger@tu-dresden.de}
\emailAdd{hyejung.stoeckinger-kim@tu-dresden.de}

\newcommand\code[1]{{\lstinline!#1!}}

\newcommand{\oldtext}[1]{\textcolor{red}{Old text removed here}}

\newcommand{\amu}{a_\mu}
\newcommand{\damu}{\Delta a_\mu}
\newcommand{\amuNEW}{\amu^{\text{2021}}}
\newcommand{\amuBNL}{\amu^{\text{BNL}}}
\newcommand{\amuFNAL}{\amu^{\text{FNAL}}}
\newcommand{\damuNEW}{\damu^{\text{2021}}}
\newcommand{\damuBNL}{\damu^{\text{BNL}}}

\FullConference{%
	*** The European Physical Society Conference on High Energy Physics (EPS-HEP2021), ***\\
	*** 26-30 July 2021 ***\\
	*** Online conference, jointly organized by Universität Hamburg and the research center DESY ***
}

\abstract{Now that the Fermilab muon $g-2$ experiment has released the results of its Run-1 data, which agrees with the results of the Brookhaven experiment, one can examine the potential of simple extensions to explain the combined $4.2\sigma$ discrepancy between the SM prediction and experiment.  This proceeding examines a single-, two-, and three-field extension of the standard model and examines their ability to explain the muon $g-2$ anomaly, and where possible, produce a dark matter candidate particle with the observed relic density.  This is based on work carried out for Ref.\ \cite{Athron:2021iuf}.  It is found that one can only explain the $\amu$ discrepancy whilst avoiding dark matter and collider constraints when the contributions from BSM fields benefit from a chirality flip enhancement.  However, in general without small couplings and/or large masses, these models can be heavily constrained by collider and dark matter experiments.  
}

\begin{document}
\maketitle
	
\section{Introduction} \label{sec:Introduction}

Recently experimentalists at Fermilab \cite{PhysRevLett.126.141801} published the first results of their measurement of the anomalous magnetic moment of the muon $\amuFNAL$, which when combined with the experimental result $\amuBNL$ from Brookhaven \cite{Bennett:2006fi} is:
\begin{align}
	\label{eqn:amuExp} 
	\amuNEW = (116592061\pm41)\times10^{-11}.
\end{align}
This differs with the prediction for $\amu$ from the Muon $g-2$ Theory Initiative  in the standard model (SM) of $\amu^{\text{SM}} = (116591810\pm43)\times10^{-11}$ \cite{aoyama2020anomalous} by $\damuNEW = (251\pm59)\times10^{-11}$, about $4.2\sigma$.  
While this is not quite large enough to confirm the existence of new physics beyond the SM (BSM) causing the discrepancy, it confirms there is tension between the SM and experiment.

There has been an abundance of analysis produced providing BSM explanations of the $\damuNEW$ discrepancy.  For a thorough review of the community's attempt to explain the $\damu$ discrepancy see Ref.\ \cite{Athron:2021iuf}.  
In addition to supersymmetric explanations of $\damu$ using the MSSM, single and two field extensions of the SM were thoroughly examined, as well as several three field models, coupling fields directly to the muon.  
In this proceeding, we focus on three BSM models from Ref.\ \cite{Athron:2021iuf} which are able to explain $\amuNEW$.  These models are simple extensions of the SM, coupling one, two, or three fields to the muon to provide contributions to $\amu$ at the first loop order.  In this proceeding we show the parameter space of each of these three models which can still explain the $\amuNEW$ discrepancy, whilst avoiding collider and other constraints.  Preferably we aim to provide a DM candidate particle with the correct abundance where possible.  

%

\section{Models} \label{sec:Models}

The scalar leptoquark we examine in this proceeding is $S_1$.  Here it is defined to be a top- and muon-philic leptoquark, coupling only to the 3rd and 2nd generations of SM quarks and leptons:
\begin{equation} \label{eqn:ScalarLeptoquarkLagrangian}
    {\cal L}_{LQ} = - (\lambda_{QL} Q_3 \cdot L_2 S_1 + \lambda_{t\mu} t \mu S_1^* + h.c.) - M_{S_1}^2 |S_1^2|,
\end{equation}
where $Q_3=(t_L,b_L)^T$, $L_2=(\nu_{\mu L},\mu_L)^T$, $t=t_R^\dagger$, $\mu=\mu_R^\dagger$.  
With these couplings to the SM, the dominant $\damu$ contributions involve a chirality flip enhancement $m_t/m_\mu$, and they simplify to (the loop functions $C$ and $F$ are defined in Ref.\ \cite{Athron:2021iuf}):  
\begin{equation} \label{eqn:ScalarLeptoquarkContribution}
    \damu^{\text{LQ}} \approx \frac{m_\mu m_t \lambda_{QL}\lambda_{t\mu}}{16\pi^2M_{S_1}^2} \begin{pmatrix} 2 Q_t F\bigg(\frac{m_t^2}{M_{S_1}^2}\bigg) - Q_{S_1} C\bigg(\frac{m_t^2}{M_{S_1}^2}\bigg) \end{pmatrix}.  
\end{equation}

Of the simple BSM models in Ref.\ \cite{Athron:2021iuf} with two fields of different spin which can provide contributions to $\amu$ whilst avoiding collider constraints, this proceeding includes the below extension:
\begin{equation} \label{eqn:Min2FieldsLL}
	{\cal L}_{2} = (\lambda_L L_2.\psi_d \phi - M_\psi \psi_d^c \psi_d + h.c.) - \frac{M_\phi^2}{2} \phi^2,
\end{equation}
which involves a scalar singlet $\phi$ paired with a $SU(2)_L$ fermion doublet $\psi_d$.  Under an introduced $Z_2$ symmetry the fields $\psi_d$ and $\phi$ are odd while the SM fields are even, and the DM candidate particle is the scalar singlet.  
The contribution to $\amu$ from these couplings is given by:
\begin{equation} \label{eqn:Min3FieldsLLContrib}
    \damu^{\text{2}} = - Q_\psi \frac{\lambda_L^2}{32\pi^2}\frac{m_\mu^2}{6M_\phi^2} E\bigg(\frac{M_\psi^2}{M_\phi^2}\bigg).  
\end{equation}
Like contributions from other models with two particles of different spin, this contribution does not get a chirality flip enhancement (of the form $M_\psi/M_\phi$ or any other), which will impact the ability of the model to explain $\damu$ later.  

The three field extension of the SM that we examine involves coupling a scalar $SU(2)_L$ singlet $\phi^0_s$, scalar doublet $\phi_d$, and a charged fermion singlet $\psi_s$, with the following couplings to the SM:
\begin{equation} \label{eqn:Min3FieldsSLRFR}
	{\cal L}_{3} = (a_H H.\phi_d \phi^0_s + \lambda_L \phi_d.L \psi_s + \lambda_R \phi^0_s \mu \psi_s^c - M_\psi \psi_s^c \psi_s + h.c.) - M_{\phi_d}^2 |\phi_d|^2 - \frac{M_{\phi_s}^2}{2} |\phi^0_s|^2.
\end{equation}
Just like the two field model, all SM/BSM fields are even/odd under a $Z_2$ symmetry.  After electroweak symmetry-breaking, the scalar singlet mixes with the neutral component of the doublet $\phi^0_s, \phi^0_d \rightarrow \phi^0_1, \phi^0_2$, 
where $\phi^0_1$ is the DM candidate particle.  
The dominant BSM contributions to $\amu$ in this model again involve a chirality flip enhancement ($\tilde{F}_a$ is defined in Ref.\ \cite{Athron:2021iuf}):
\begin{equation} \label{eqn:Min3FieldsSLRFRContrib}
	\damu^{\text{3}} \approx - \frac{m_\mu^2}{32\pi^2M_\psi^2} \frac{2\sqrt{2}\lambda_L\lambda_R a_H v}{m_\mu M_\psi} \tilde{F}_a\bigg(\frac{M_\psi^2}{M_{\phi^0_1}^2},\frac{M_\psi^2}{M_{\phi^0_2}^2}\bigg).
\end{equation}






\section{Results} \label{sec:Results}

\begin{figure}[ht!]
    	\centering
   		\includegraphics[width=0.32\textwidth]{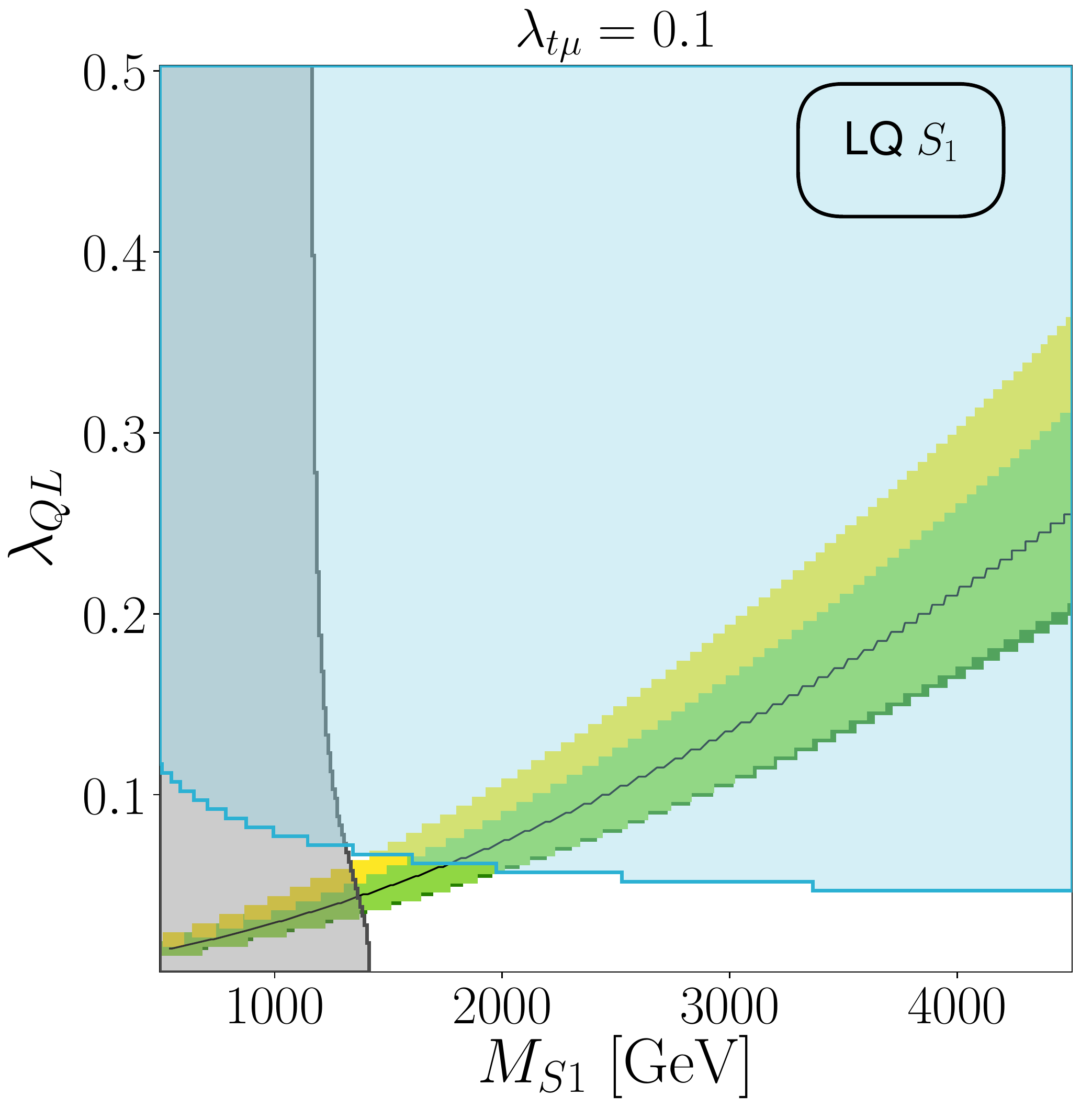}
   		\includegraphics[width=0.32\textwidth]{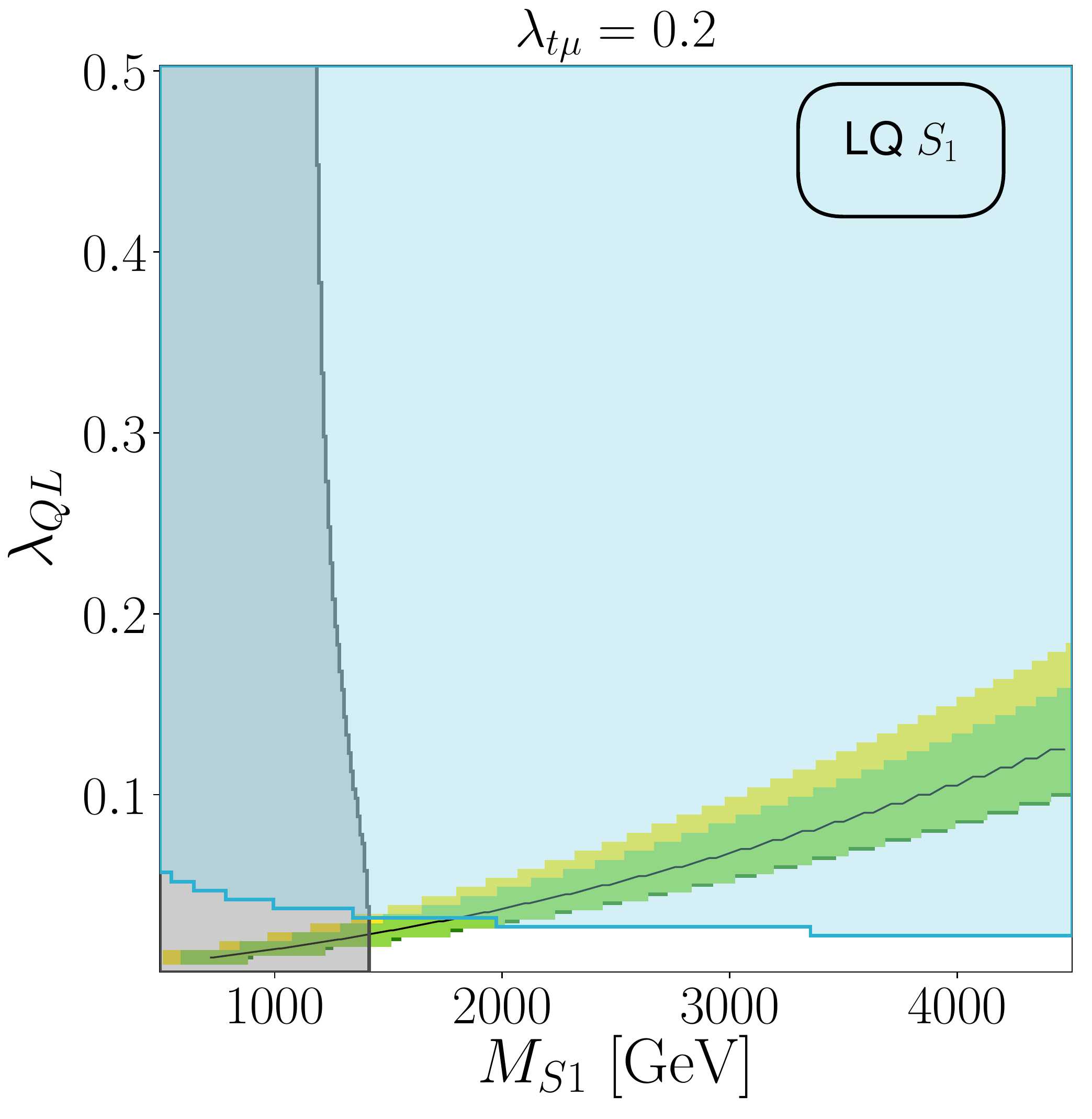}
   		\includegraphics[width=0.32\textwidth]{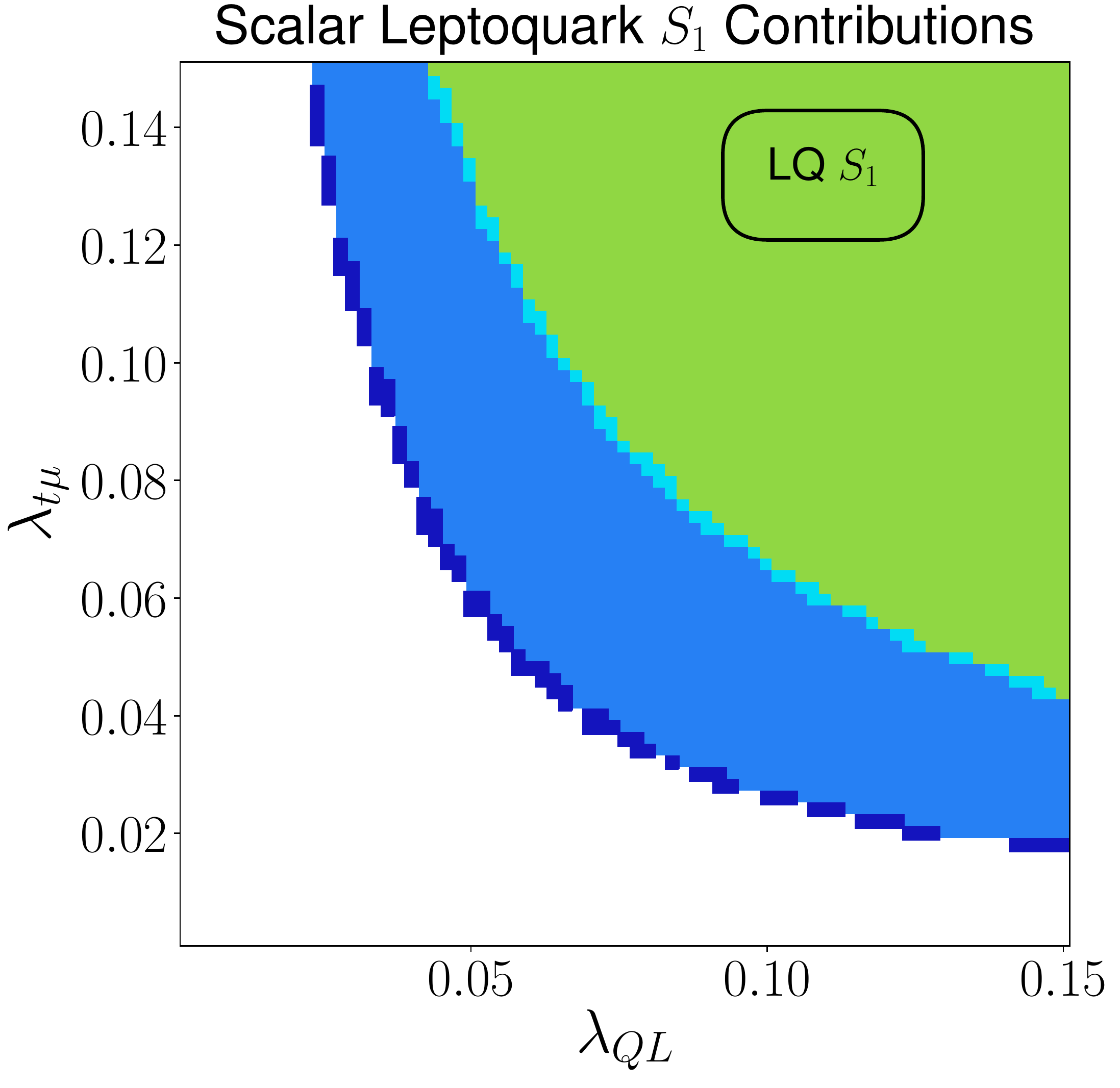}
    \caption{Contributions to $\amu$ from the leptoquark $S_1$ defined in Eq.\ (\ref{eqn:ScalarLeptoquarkLagrangian}).  The yellow and green regions indicate the points which can explain the discrepancies $\damuBNL$ and $\damuNEW$ from Eq.\ (\ref{eqn:amuExp}) within $1\sigma$, while the lime green region indicates the overlap.  The black line indicates points which provide a contribution which exactly explains $\amuNEW$.  Points constrained by LHC searches for leptoquarks are shaded grey \cite{Sirunyan:2018ruf}.  The left panels have regions shaded cyan indicating points disfavoured by muon mass fine-tuning considerations.  The right panel shows a profile over the leptoquark mass $M_{LQ}$, and points which can explain the $\damuBNL$, $\damuNEW$ discrepancies and their overlap whilst avoiding fine-tuning are shown in cyan, dark blue and blue respectively.  \label{fig:ScalarLeptoquarkSinglet} }
    \end{figure}

The contributions to $\amu$ from the leptoquark $S_1$ defined in Eq.\ (\ref{eqn:ScalarLeptoquarkLagrangian}) are shown in Fig.\ \ref{fig:ScalarLeptoquarkSinglet}, which are calculated using FlexibleSUSY \cite{Athron2018}.  In the left and middle panels we have fixed a coupling to $\lambda_{t\mu}=0.1$ and $0.2$.  Then the points which can explain $\damuNEW$ have a quadratic relationship between $\lambda_{QL}$ and $M_{LQ}$, as indicated by Eq.\ (\ref{eqn:ScalarLeptoquarkContribution}).  However, there are constraints on the parameter space from LHC leptoquark searches \cite{Sirunyan:2018ruf}, placing a rough lower limit of $M_{LQ} \gtrsim 1400$ GeV.  Furthermore, points with higher couplings are disfavoured since they give a contribution to the muon mass which shifts it outside the bounds of $m_\mu\in[m_{\mu,\text{pole}}/2,2m_{\mu,\text{pole}}]$, requiring fine-tuning.  
The right panel of Fig.\ \ref{fig:ScalarLeptoquarkSinglet} shows a profile over the leptoquark's mass $M_{LQ}$, excluding points which violate LHC constraints.   Overall, explaining $\damuNEW$ whilst avoiding LHC constraints requires $\lambda_{QL}\times\lambda_{t\mu} \gtrsim 0.003$, but to avoid having a fine-tuned $m_\mu$, we get an upper bound on the couplings. 

\begin{figure}[ht!]
    	\centering
   		\includegraphics[width=0.32\textwidth]{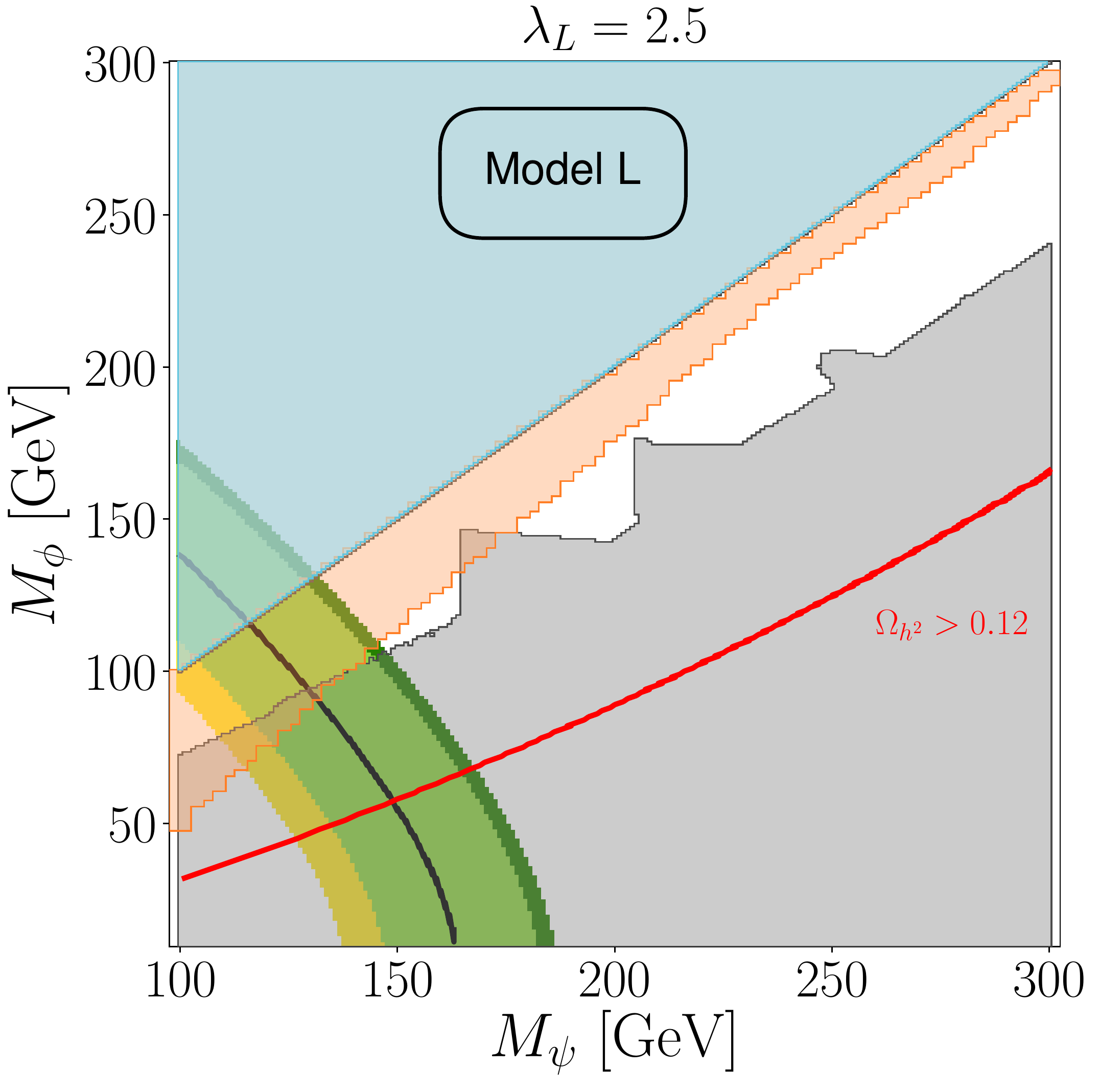}
   		\includegraphics[width=0.32\textwidth]{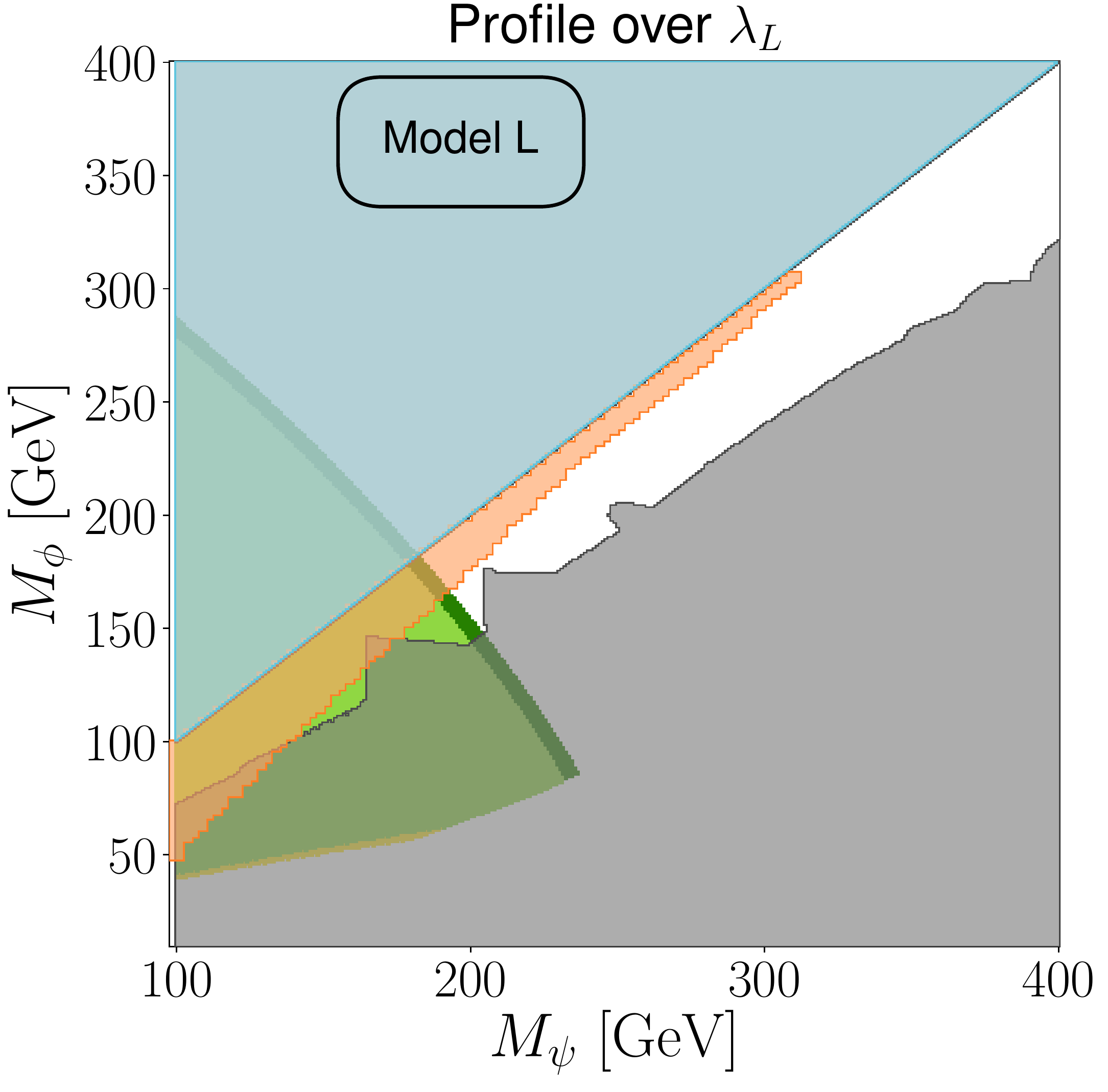}
   		\includegraphics[width=0.32\textwidth]{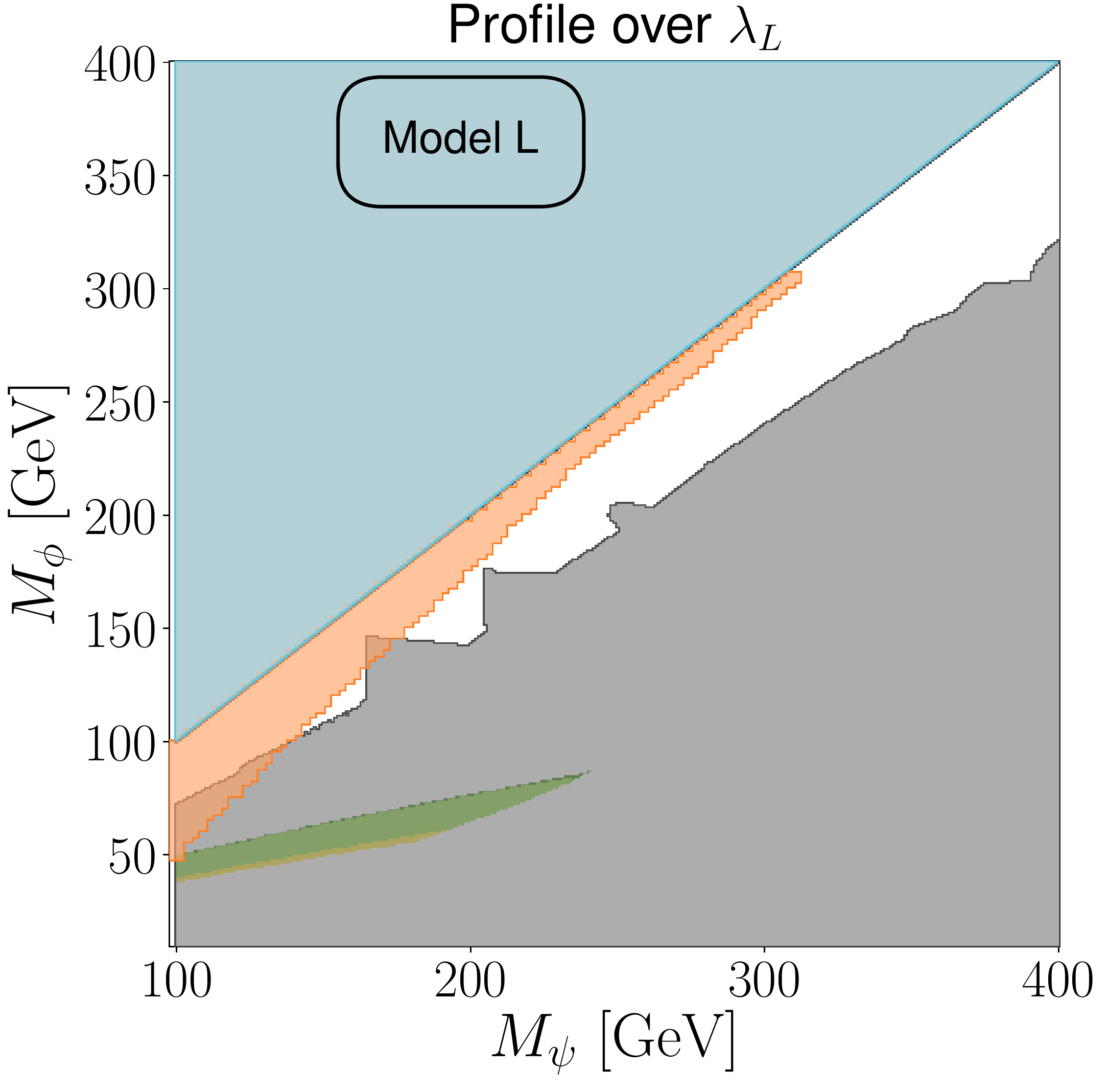}
    \caption{Contributions to $\amu$ from extending the SM with two fields defined in Eq.\ (\ref{eqn:Min2FieldsLL}).  The colours for $\amu$ contributions are the same as in Fig.\ \ref{fig:ScalarLeptoquarkSinglet}.  The red line indicates masses which produce a DM relic density matching the Planck observation in Ref.\ \cite{Aghanim:2018eyx} calculated using micrOMEGAs 5.2.1 \cite{Belanger2018}.  Points above this line have an under abundant relic density, while points under the line are over abundant and strongly excluded.  Constraints from the LHC searchs in Refs.\ \cite{Aad:2014vma,Sirunyan:2018nwe} calculated using SModelS 1.2.3 \cite{Khosa2020} are indicated by the region shaded grey, while compressed spectra search constraints from Ref.\ \cite{Sirunyan:2018iwl} calculated using CheckMATE 2.0.26 \cite{Dercks2017} are shaded orange.  The region with charged DM is shaded cyan.  The middle and right panels profile over $\lambda_L$, where the middle panel excludes points with over abundant relic density, and the right panel only shows points with relic density within $1\sigma$ of the Planck observation.  \label{fig:Min2FieldsLL} }
    \end{figure}

The contributions to $\amu$ from the two fields extension of the SM defined in Eq.\ (\ref{eqn:Min2FieldsLL}) are shown in Fig.\ \ref{fig:Min2FieldsLL}.  In the left panel, we fixed $\lambda_L=2.5$ and scanned over the masses $M_\psi$ and $M_\phi$.  We can explain the $\damuNEW$ discrepancy in the regions curved in a parabola, following the ratio relationship between the BSM masses shown in Eq.\ (\ref{eqn:Min3FieldsLLContrib}).  However, these regions are almost completely excluded by constraints from the LHC searches in Ref.\ \cite{Aad:2014vma,Sirunyan:2018nwe} and Ref.\ \cite{Sirunyan:2018iwl}.  

Futhermore, the strong LHC constraints exclude a simultaneous explanation of $\damuNEW$ and DM.  The masses which produce a DM candidate particle with the Planck-observed relic density of $\Omega_{h^2} = 0.1200\pm0.0020$ \cite{Aghanim:2018eyx} are indicated by the red line.  In the left panel of Fig.\ \ref{fig:Min2FieldsLL}, the LHC constraints rule out all the masses with the observed DM relic density.  If we profile over the couplings $\lambda_L$ between the BSM fields and the muon in the middle panel using a targeted MultiNest 3.10 scan \cite{Feroz:2013hea}, then we can explain $\damuNEW$ in two small regions of parameter space between the LHC searches, making future compressed spectra searches important for fully excluding this model.  However, this only requires that the DM candidate particle not be over abundant.  If we require that the relic density matches the Planck observation, as in the right panel, then the masses produce a contribution which explains $\damuNEW$ is excluded by LHC constraints.  This arises because the contribution to $\amu$ in Eq.\ (\ref{eqn:Min3FieldsLLContrib}) has no chirality flip enhancement, requiring small masses to explain the anomaly, putting it into conflict with collider constraints.  

\begin{figure}[ht!]
    	\centering
   		\includegraphics[width=0.32\textwidth]{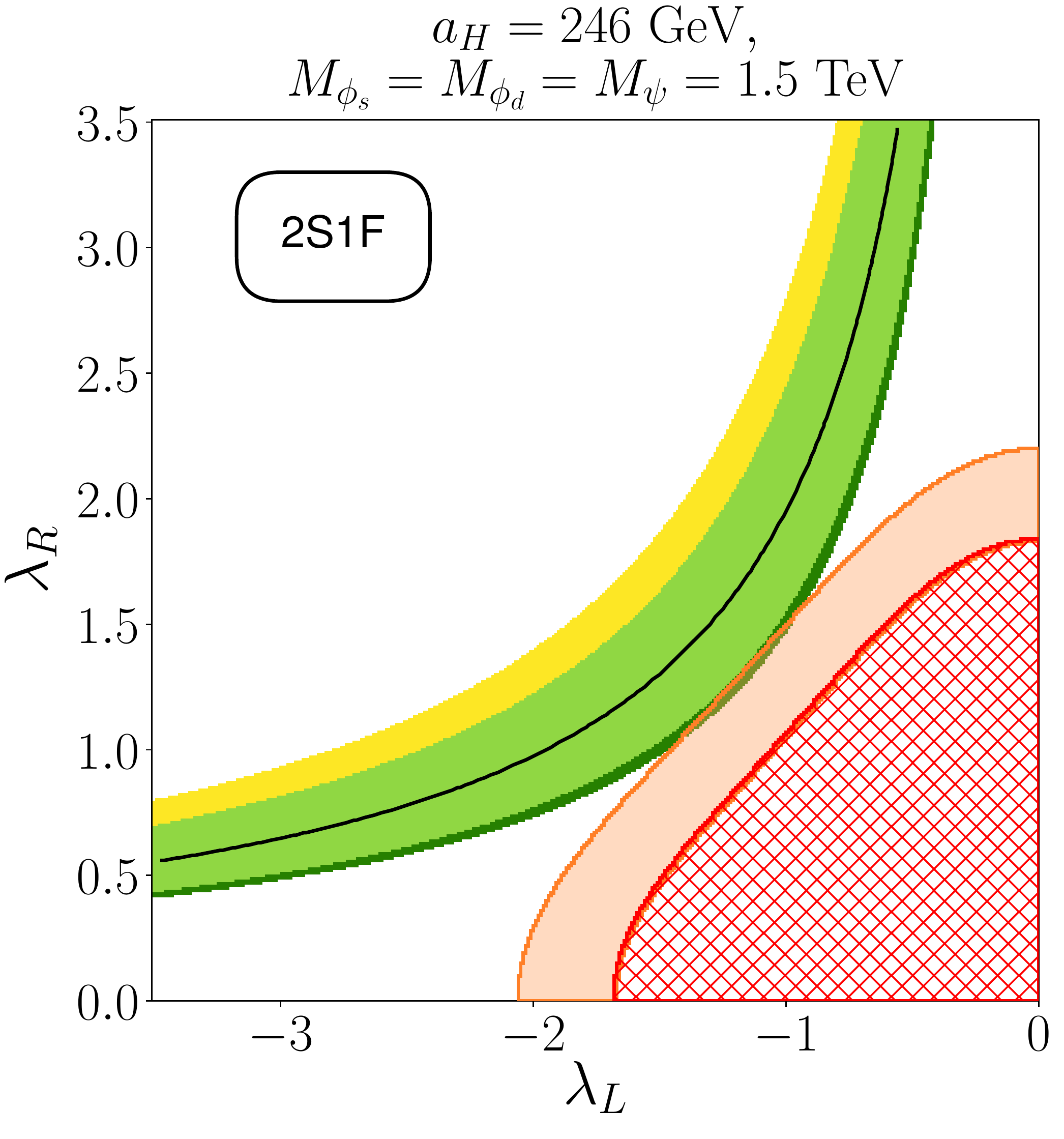}
   		\includegraphics[width=0.32\textwidth]{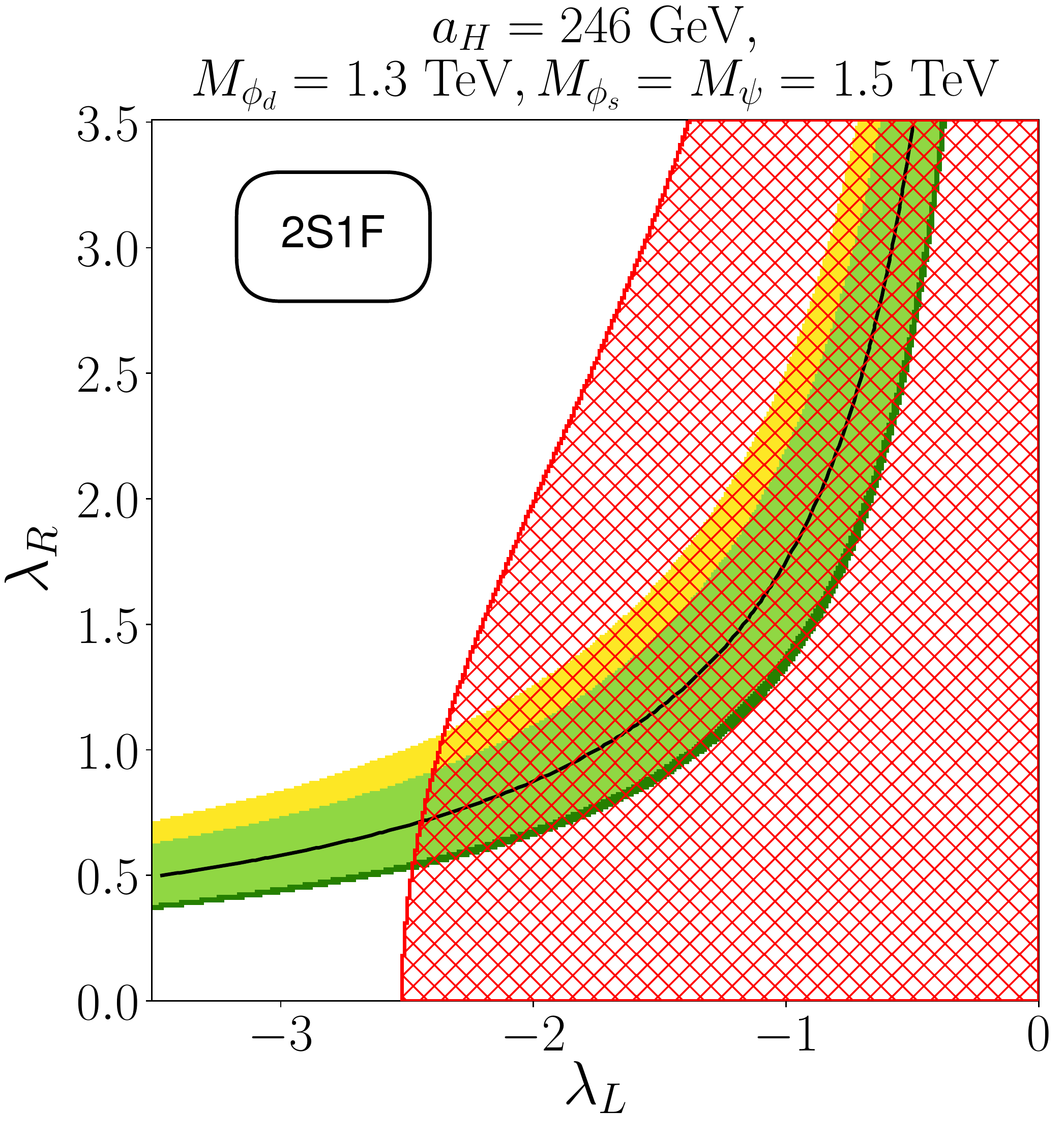}
   		\includegraphics[width=0.32\textwidth]{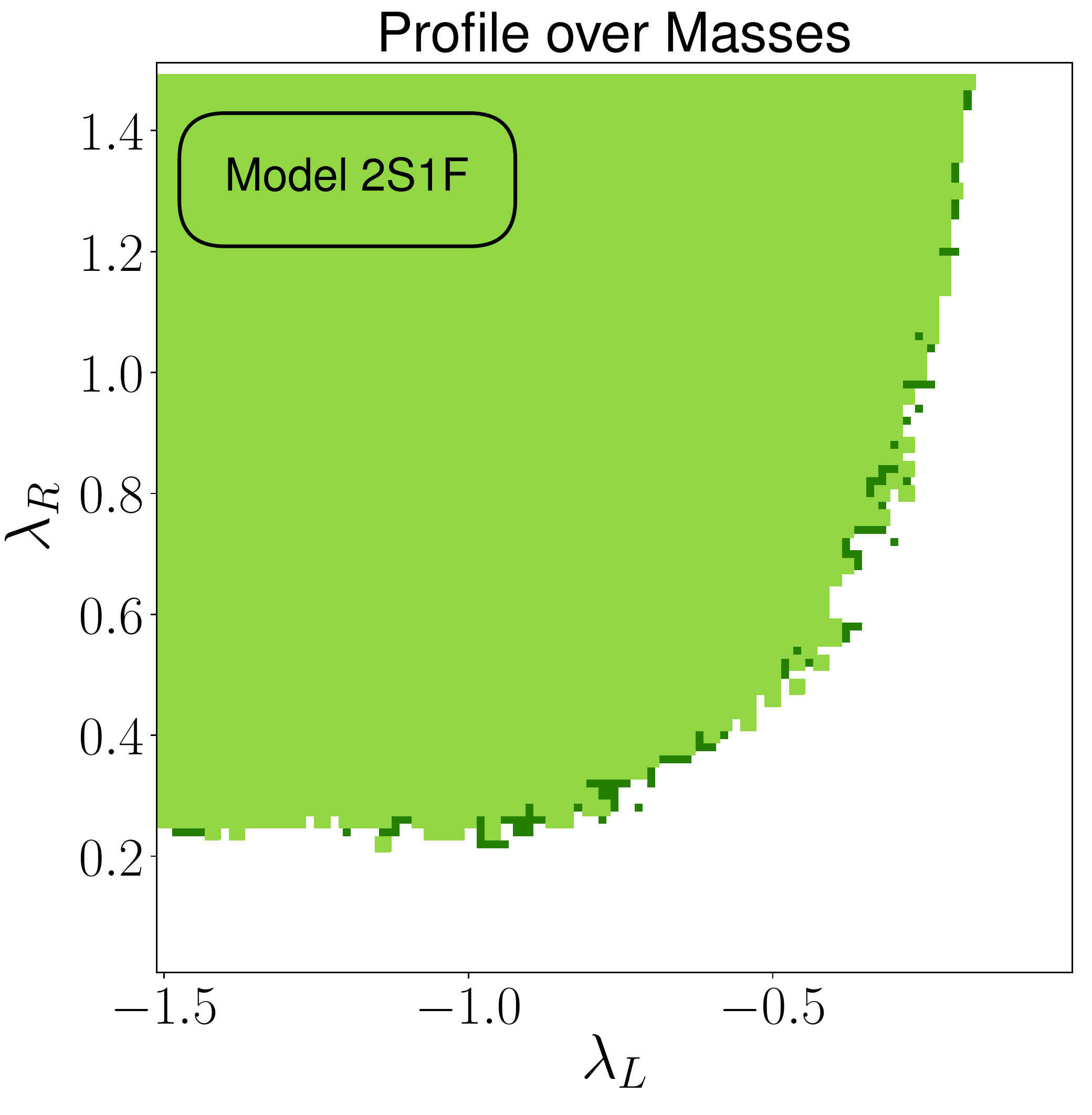}
    \caption{Contributions to $\amu$ from extending the SM with three fields defined in Eq.\ (\ref{eqn:Min3FieldsSLRFR}).  The colours for $\amu$ contributions are the same as in Fig.\ \ref{fig:ScalarLeptoquarkSinglet}.  In the left panels, points with an over abundant DM relic density are indicated with the red hatched region, while the region constrained by direct detection searches at the Xenon1T experiment (see Ref.\ \cite{Aprile:2018dbl}) is shaded orange.  The right panel is a profile over all the BSM masses and the coupling $a_H$, where points which are ruled out by direct detection searches or have a relic density more than $3\sigma$ from the Planck observation are excluded.  \label{fig:Min3FieldsSLRFR} }
    \end{figure}

The three fields model defined in Eq.\ (\ref{eqn:Min3FieldsSLRFR}) does however have a chirality flip enhancement for its contributions to $\amu$ in Eq.\ (\ref{eqn:Min3FieldsSLRFRContrib}), which depends on the product of the couplings $\lambda_L\times\lambda_R$.  The left panel of Fig.\ \ref{fig:Min3FieldsSLRFR} shows the scenario where all of the BSM fields have equal masses.  Due to the large masses of the BSM fields, we can avoid constraints from collider searches.  However, since the BSM masses are all set to the same value, $1.5$ TeV, almost all possible DM depletion mechanisms are turned on, leading to quite large constraints from the Xenon1T \cite{Aprile:2018dbl} direct detection experiment (calculated through DDCalc 2.2.0 \cite{Athron:2018hpc}).  Points with smaller couplings are excluded by an over abundant relic density.  We can turn off many of these DM depletion mechanisms by splitting the masses, as shown in the middle panel, where the scalar doublet $\phi_d$ is made $200$ GeV lighter than the other BSM masses.  Then the DM candidate $\phi_1^0$ becomes doublet-dominated, and singlet depletion mechanisms depending on $\lambda_R$ weaken, leading to the DM relic density depending much more on $\lambda_L$, while the direct detection constraints weaken.  
	
The right panel of Fig.\ \ref{fig:Min3FieldsSLRFR} shows a profile over the values $M \in [0,5000]$ GeV, $a_H \in [0,5000]$ GeV, and $|\lambda_{L,R}| \in[0,1.5]$ using a MultiNest scan.  Points which violated direct detection constraints, or produced a DM candidate with a relic density that did not match the Planck observation were discarded.  It was found that the product of couplings needed to be at least $|\lambda_L\times\lambda_R| \gtrsim 0.22$ to explain $\damuNEW$ and DM simultaneously.  This lower limit comes from the depletion of doublet-dominated DM from $SU(2)_L$ interactions, and the inability of singlet-dominated DM to explain the $\amu$ anomaly and DM simultaneously.  For DM which is a strong mixture of the singlet and doublet, the lower bound is set through depletion via $a_H$ and strengthened direct detection constraints.  

\section{Conclusions} \label{sec:Conclusions}

The latest results from the Fermilab $\amu$ experiment have reconfirmed the discrepancy between the SM and experiment, suggesting BSM contributions.  There are many simple extensions of the SM, and we have looked into a few which can explain the $\damuNEW$.  With additional higher precision experimental results to be released in the future, more simple BSM models will be unable to explain $\amu$, and the bounds on the parameters of those which can still explain the discrepancy will tighten.  

\bibliographystyle{ieeetr}
\bibliography{TheoryWPbiblio,TheWork,DouglasJacobDarkMatter,DouglasJacobMuonMoment,DouglasJacobMuonMomentExperiment,DouglasJacobMuonMomentLeptoquarks,DouglasJacobPackages,DouglasJacobATLAS,DouglasJacobCMS,DouglasJacobMuonMoment2HDM,CodeReferences,SModelSdatabasereferences,SectionSUSYbiblio}

%
	
\end{document}